\documentclass[final,5p,times,twocolumn]{elsarticle}

\usepackage{amssymb}
\usepackage{amsthm}
\usepackage{amsmath}
\usepackage{float}
\usepackage{subcaption}
\usepackage{booktabs}
\usepackage{adjustbox}

\usepackage[pagewise]{lineno}
\modulolinenumbers[5]
\usepackage{hyperref}
\usepackage{cleveref}
\journal{Fuel Communications}

\begin{document}

\begin{frontmatter}



\title{Intrinsic instability of lean hydrogen/ammonia premixed flames: Influence of Soret effect and pressure}

\author[inst1]{F. D'Alessio*}
\author[inst1]{P.E. Lapenna}
\author[inst1]{F. Creta}

\affiliation[inst1]{organization={Mechanical and Aerospace Engineering, Sapienza University of Rome},
            addressline={Via Eudossiana 18}, 
            city={Rome},
            postcode={00186}, 
            country={Italy}}

\begin{abstract}
The addition of hydrogen in ammonia/air mixtures can lead to the onset of intrinsic flame instabilities at conditions of technical relevance. The length and time scales of intrinsic instabilities can be estimated by means of linear stability analysis of planar premixed flames by evaluating the dispersion relation. In this work, we perform such linear stability analysis for hydrogen-enriched ammonia/air flames (50\%H2-50\%NH3 by volume) using direct numerical simulation with a detailed chemical kinetic mechanism. The impact of pressure and the inclusion of the Soret effect in the governing equations is assessed by comparing the resulting dispersion relation at atmospheric pressure and 10 atm. Our data indicate that both pressure and the Soret effects promote the onset of intrinsic instabilities. Comparisons with available numerical literature data as well as theoretical models are also discussed.   
\end{abstract}



\end{frontmatter}



\section{Introduction}\label{sec: Intro}
Global trends indicate that combustion will remain a key technology for energy conversion throughout the remainder of the twenty-first century~\cite{conti2016}. To achieve a sustainable combustion process, research is focusing on carbon-free fuels such as Ammonia (NH3) and Hydrogen (H2) which do not emit any carbon oxides (CO, CO2) when burned with air~\cite{dreizler2021role}. However, their implementation in industrial devices remains economically and technically challenging. In this context, hydrogen offers attractive properties as an energy carrier, but its combustion in power plants raises significant safety concerns~\cite{Abohamzeh2021}. Its high reactivity and peculiar thermochemical and transport properties~\cite{lapenna2023hydrogen}, compared to hydrocarbon fuels, leads to significant design barriers in terms of flame stability, unpredictable ignition of leaks, transition to detonation and ultimately nitrogen oxide (NOx) emissions. In addition, hydrogen production, storage, and transportation remain significantly uneconomical for large-scale industrial use~\cite{tingas2023hydrogen}. On the other hand, ammonia is emerging as an efficient hydrogen carrier.
From an economic standpoint, the production, transportation, and storage of ammonia are significantly simpler and more cost-effective than that of hydrogen~\cite{Dincer2018}. Although direct combustion of ammonia has also been considered as an alternative to hydrogen, its low reactivity and high emissions of NOx represent significant challenges to its direct use. A possible approach to overcome these issues is to blend H2 and NH3 to leverage the properties of both fuels~\citep{Elbaz2022, Tornatore2022}. 

The addition of hydrogen in a fuel blend can promote the onset of Intrinsic Flame Instabilities (IFIs) at conditions and scales of technical interest. IFIs affect the characteristics of the flame, the amount of heat released, and the overall flame morphology and propagation. Unlike thermoacoustic instabilities, IFIs are endogenous to the flame and not linked to the geometry of the combustor and the ensuing pressure waves or oscillations~\cite{Matalon2007}. The two main mechanisms at play in the onset of IFIs are the hydrodynamic or Darrieus-Landau instability (DL) and the thermodiffusive instability (TD). The DL instability is caused by the local flow field induced by the density gradient across a premixed flame which is therefore active at all flame perturbation wavenumbers irrespective of the mixture employed~\cite{Matalon2018}. On the other hand, the TD mechanism actively destabilizes the flame when the effective Lewis number of the mixture is lower than a critical value $Le_0$, which is typically close to unity~\cite{Creta2020}. Given a slightly perturbed (curved) flame front, this introduces a disparity between transverse heat and reactant fluxes, which in turn gives rise to enhanced and reduced reaction zones that amplify the perturbation~\cite{lapenna2023hydrogen}. In the presence of molecular and atomic hydrogen (H2, H), the disparity between molecular and thermal diffusivity is large enough to cause significant local modification of the flame speed, which results, in the non-linear regime, in a cellular structure~\cite{Frouzakis2015}. Early and extensive asymptotic studies have been carried out on the stability of planar flames such as the work of~\citet{matalon_matkowsky_1982},\citet{clavin_williams_1982} and~\citet{Sivashinsky}. While extremely useful for a qualitative characterization, the ensuing theoretical models for the description of the flame stability are generally not suitable for a quantitative description of realistic mixtures, in particular when hydrogen is present and the TD instability mechanism is active. Therefore, to characterize the stability of the hydrogen-enriched ammonia flames of interest for this work, Direct Numerical Simulations (DNS) are utilized, featuring an accurate description of the thermochemical and transport properties of the mixture.  

Numerous numerical studies based on DNS have investigated the onset and impact of IFIs~\cite{kadowaki2005unstable,kadowaki2005_numerical,yuan2007flame,sharpe2006nonlinear} while more recently, the linear stability analysis of perturbed planar premixed flames has been investigated through DNS for various mixtures and conditions. The effect of pressure on hydrodynamic instabilities has been investigated by~\citet{Attili2021} and~\citet{lamioni2020pressure} for lean methane-air flames. On the other hand, for TD unstable mixtures, such as lean hydrogen-air flames, the numerical dispersion relation has been evaluated and compared to theoretical results by~\citet{altantzis2012hydrodynamic}. Berger~et~al.~\citep{Berger2019,Berger2022_1} conducted a comprehensive investigation to determine the numerical dispersion relation of lean hydrogen-air flames spanning several parameters, including pressure, initial temperature, and equivalence ratio. They found that IFIs are promoted using leaner mixtures and increasing the pressure, while keeping the fresh gas temperature low. However, significantly fewer DNS studies are currently available for hydrogen-enriched ammonia flames. Preferential diffusion effects were observed by~\citet{Rieth2022} in a partially cracked mixture of ammonia, hydrogen, and nitrogen in air. Similarly, \citet{Wiseman2021} observed an abrupt increase in fuel consumption and flame surface area in lean premixed turbulent ammonia-air flame under partially cracked conditions. \citet{Rieth2023} and~\citet{Netzer2021} explored the impact of IFIs and flame topology on NOx formation in a turbulent premixed NH3/H2/air flame under varying equivalence ratios and pressures, finding a correlation between topology changes caused by hydrogen preferential diffusion and NOx generation. A parametric analysis of the stability limits of a premixed flame of ammonia, hydrogen, and air has been recently conducted by~\citet{Gaucherand2023} showing how the growth rate of perturbations is modified by the equivalence ratio, composition, and pressure. However, they employed a simplified transport model without the inclusion of thermophoresis (Soret effect) which, in flames featuring large amounts of hydrogen can have a significant role~\cite{Schlup2018_2}.  

In this framework, we perform linear stability analysis of planar hydrogen-enriched
ammonia/air flames (50\%H2-50\%NH3 by volume) using DNS with detailed chemical kinetic mechanism and transport models. The objective is to evaluate the impact of pressure and the role of the Soret effect on the dispersion relations and the resulting flame stability limits. The latter is of fundamental importance to estimate the IFI's length and time scales as they have a significant impact on the prediction of flame probation. Such values are indeed needed by newly developed combustion models that are accounting for IFIs at the subgrid level~\cite{Lapenna2021_subgrid,LAPENNA2024}. 

\section{Theoretical an numerical framework}\label{sec: Method}
\subsection{Governing equations}
In the following, we assume low-Mach number conditions, which are standard for deflagrating fronts, as well as detailed chemical kinetics. All mixture properties are calculated using mixture-averaged formulations based on pure species properties~\cite {kee1986fortran}. Diffusion coefficients are computed with the Hirschfelder-Curtiss approximation and a velocity correction $\mathbf{V}_c$ is introduced to enforce mass conservation both in the $N_s$ species and temperature equations. To decrease the computational cost, the Soret effect is accounted for only for lighter species (H2, H) following the formulation and parameters introduced by Schlup and Blanquart~\citep{Schlup2018_1,Schlup2018_2}. Under these assumptions, the diffusion velocity reads:

\begin{equation}
    \mathbf{V}_i=-D_i \frac{\nabla Y_i}{Y_i}-D_i\frac{\nabla W}{W}-\frac{D_i^T}{\rho Y_i} \frac{\nabla T}{T} +\mathbf{V}_c
\end{equation}
\noindent where the correction velocity is
\begin{equation}
    \mathbf{V}_c=\textstyle\sum_k^{N_s} D_k \nabla Y_k + \frac{\nabla W}{W} \textstyle\sum_k^{N_s} D_k Y_k + \frac{1}{\rho}\frac{\nabla T}{T}\textstyle\sum_k^{N_s} D_k^T
\end{equation}

\noindent and the system of governing equations reads as follows:\\

\noindent \textbf{Continuity equation} in terms of thermal divergence:
\begin{equation}
\nabla \cdot  \mathbf{u} = - \frac{1}{\rho} \frac{D \rho}{Dt}=\sum_k^{Ns} \frac{W}{W_k}\frac{D Y_k}{Dt}+\frac{1}{T}\frac{D T}{Dt}
\end{equation}
\textbf{Momentum equation}:    
\begin{equation}
\rho \frac{D\mathbf{u}}{Dt} = -\nabla p_1 + \nabla \cdot 
\left( \mu\left[\nabla \mathbf{u} + (\nabla \mathbf{u})^T - \frac{2}{3}(\nabla \cdot \mathbf{u})\right]\mathbf{I}\right) 
\end{equation}
\textbf{Species mass fraction equation}:             
\begin{multline}
\rho\frac{D Y_i}{Dt} = \nabla \cdot ( \rho D_i \nabla Y_i)  - \nabla \cdot \Bigr(\rho Y_i \textstyle\sum_k^{N_s} D_k \nabla Y_k\Bigr) + \\ - \nabla \cdot ( F_i \nabla W) - \nabla \cdot ( H_i \nabla T) + \dot{\omega}_i
\label{spec_eq}
\end{multline}
\textbf{Temperature equation}:
\begin{multline}
      \rho C_p \frac{D T}{Dt} = \nabla \cdot ( \lambda \nabla T) - \Bigr( \textstyle\sum_k^{N_s} P_k \nabla Y_k \Bigr) \cdot \nabla T + \\ - Q \nabla W \cdot \nabla T - Q^s \nabla T \cdot \nabla T - \sum_i^{N_s} h_i \dot{\omega}_i 
      \label{temp_eq}
\end{multline}

where $\rho$ is the mixture density, $T$ the temperature, $p_0$ the thermodynamic background pressure, and where the ideal gas equation of state (EoS) reads $p_0 = \rho RT$, $R$ being the gas constant. In addition, $p_1$ is the hydrodynamic pressure, $W$ the mean molecular weight of the mixture, $Y_i$ the species' mass fractions, $C_p$ the heat capacity at constant pressure of the mixture, $\lambda$ is the mixture thermal conductivity, $D_i$ is the diffusion coefficient and $D_i^T$ the Soret thermal diffusion coefficients of the $i-$th species, $h_i$ its enthalpy and $\dot{\omega}_i$ is the $i-$th species reaction rate. The reaction rate term is established through Arrhenius kinetics and all thermochemical and transport coefficients (except for the Soret coefficients) are established through the CHEMKIN-II package~\cite{chemkin}, tabulated and stored as a function of temperature for each species, while the mixing rules~\cite{kee1986fortran} are evaluated at runtime. The coefficients in r.h.s. of the species equation are defined as
\begin{equation}
    \begin{cases}
    F_i =  \frac{\rho Y_i}{W} \left(\sum_k^{Ns}D_kY_k-D_i \right)\\    
    H_i = \frac{1}{T}\left(Y_i \sum_k D_{k}^T - D_{i}^T\right)
    \end{cases} 
\end{equation}
and the coefficients in r.h.s. of the temperature equation are defined as
\begin{equation}
    \begin{cases}
    P_k =\rho D_k \left( C_p -  C_{p,k}\right)\\    
    Q\ =\sum_{i}^{Ns} C_{p,i}F_i\\
    Q^s =\sum_{i}^{Ns} C_{p,i}H_i
    \end{cases} 
\end{equation}

This set of governing equations and models are implemented in the low-Mach number, massively parallel, spectral element~\cite{Fischer2009} flow solver Nek5000~\cite{nek5000}. This numerical framework has been developed starting from a previous version featuring one-step chemistry~\citep{Lapenna2019_1,Lapenna2019_2,Lamioni2019,lamioni2020pressure}. The code employs a high-order splitting for reacting flows~\cite{Tomboulides1998} and the chemical source terms are implicitly integrated using the stiff ODE solver CVODE~\citep{Hindmarsh2005}. The high-order characteristics of the framework are well-suited to efficiently perform combustion DNS, capturing both small-scale flame features and fast chemical time scales with minimum numerical dissipation and dispersion over extended integration times. 

\subsection{Code validation}
The code is preliminarily validated by comparing a planar flame simulation with a reference solution obtained with Cantera~\cite{cantera}. For this test, a lean methane-air mixture is used as reported in Tab.~\ref{tab: Val-flame}. The computational grid has a length of $50 \ell_T$ in the flame propagation direction, where $\ell_T = (T_b -T_u)/\nabla T_{max}$ is the thermal flame thickness, with a zero gradient condition at the outlet, while a $2 \ell_T$ width in the spanwise direction with periodic boundary conditions. 
A spatial resolution of $36 \mu m$, equivalent to $\sim20$ points in the flame thickness is used. A stationary flame front is maintained within the domain by providing a fixed velocity inlet with a velocity matching that of the laminar flame speed $S_L^0$. A total time of $50 \tau_F$ was simulated to ensure independence from the initial conditions, where $\tau_F= \ell_T/S_L^0$ is the characteristic laminar flame time. 

\begin{table}
    \centering
    \begin{adjustbox}{width=\columnwidth}
    \begin{tabular}{c c c c c c c c}
    \toprule
    Mixture &$T_{u}$~[K] & $p_0$~[Atm] & $\Phi$ & $\ell_T$~[$\mu$m] & $S_L^0$~[cm/s] & $\tau_F$~[$\mu$s] & Prog. Var. def. \\
    \midrule
    CH$_4$-Air &$300$ & $1$ & $0.70$ & $732$ & $17.4$ & $4172$ & $C = 1-\frac{Y_{O2}-Y_{O2}^b}{Y_{O2}^u-Y_{O2}^b}$\\
    \bottomrule
    \end{tabular}
    \end{adjustbox}
    \caption{Thermochemical parameters of reference validation flame.}
    \label{tab: Val-flame}
\end{table} 

\begin{figure}
    \centering
    \includegraphics[width=\columnwidth]{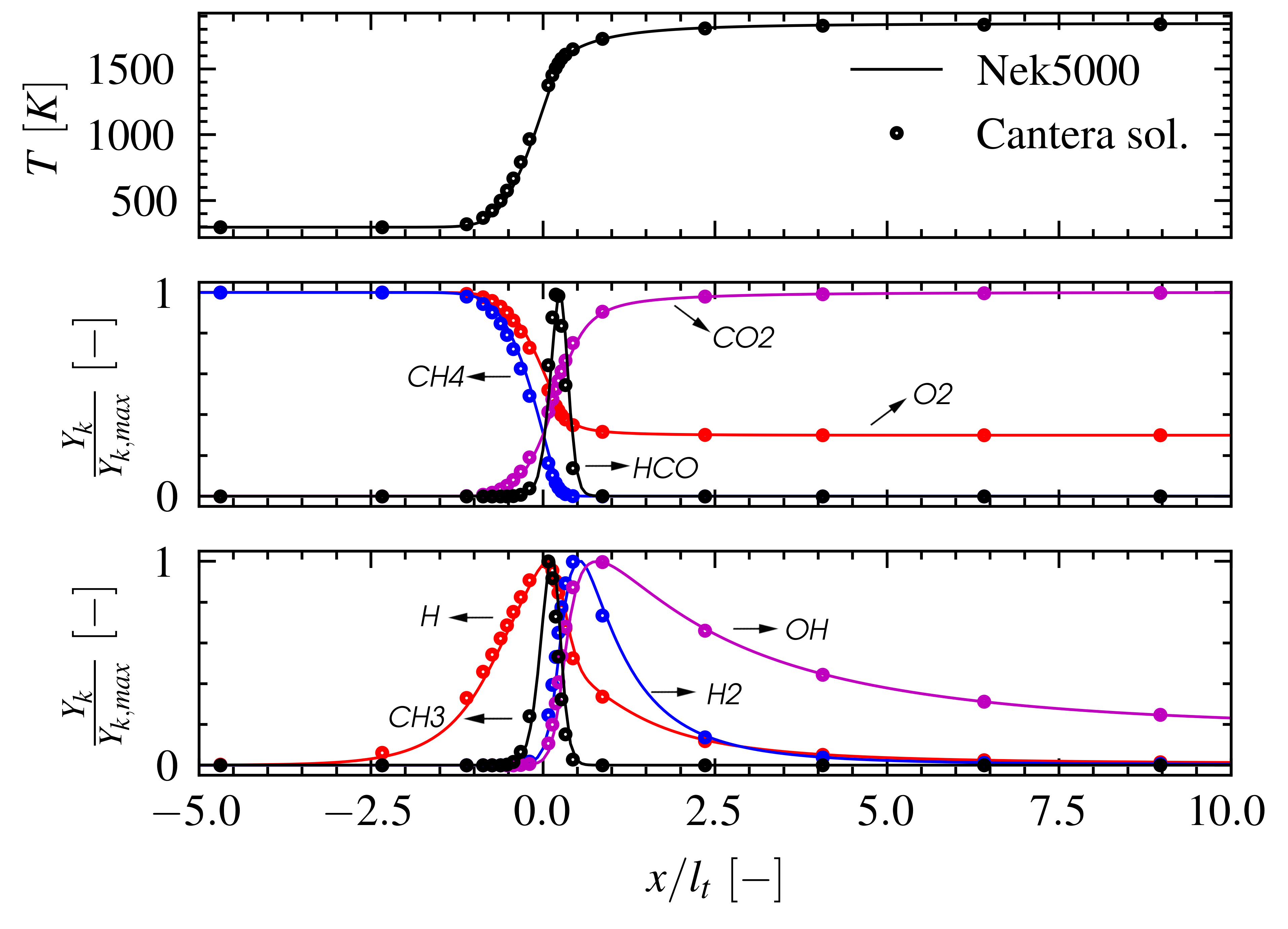}
    \caption{Comparison of species and temperature profiles in the propagation direction.}
    \label{fig: Validation}
\end{figure}

\figurename~\ref{fig: Validation} shows the comparison of species profiles obtained from the DNS code with the reference results from Cantera using, for both cases, a skeletal mechanism tailored for lean methane/air flames~\cite{Luca2018}. A good agreement is observed throughout the flame with the DNS code able to reproduce both the temperature and main species profiles as well as the radicals' sharp variation in the reacting region.  

We conducted a grid sensitivity analysis to evaluate the impact of spatial resolution on the code capability to resolve the flame structure and capture the correct flame front propagation velocity. Three grids were used with varying spatial resolutions, namely: $73 \mu m$ for the Coarse grid, $48 \mu m$ for the Medium grid, and $37 \mu m$ for the Fine grid corresponding to $\sim10$,$\sim15$ and $\sim20$ points in the flame thickness, respectively. 

\begin{figure}
    \centering
    \begin{subfigure}[b]{0.45\columnwidth}
        \includegraphics[width=\columnwidth]{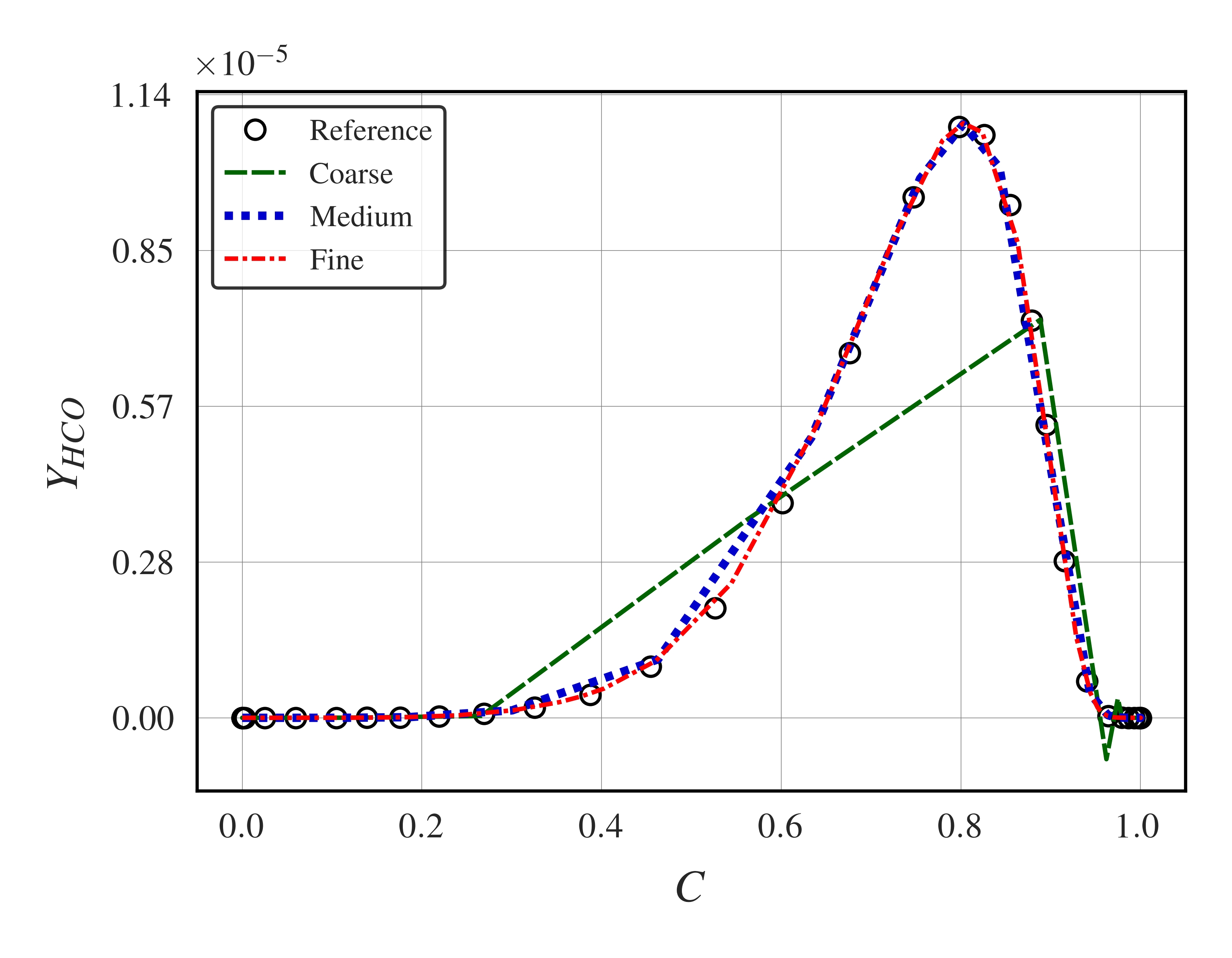}
    \end{subfigure}
    \hfill
    \begin{subfigure}[b]{0.45\columnwidth}
        \includegraphics[width=\columnwidth]{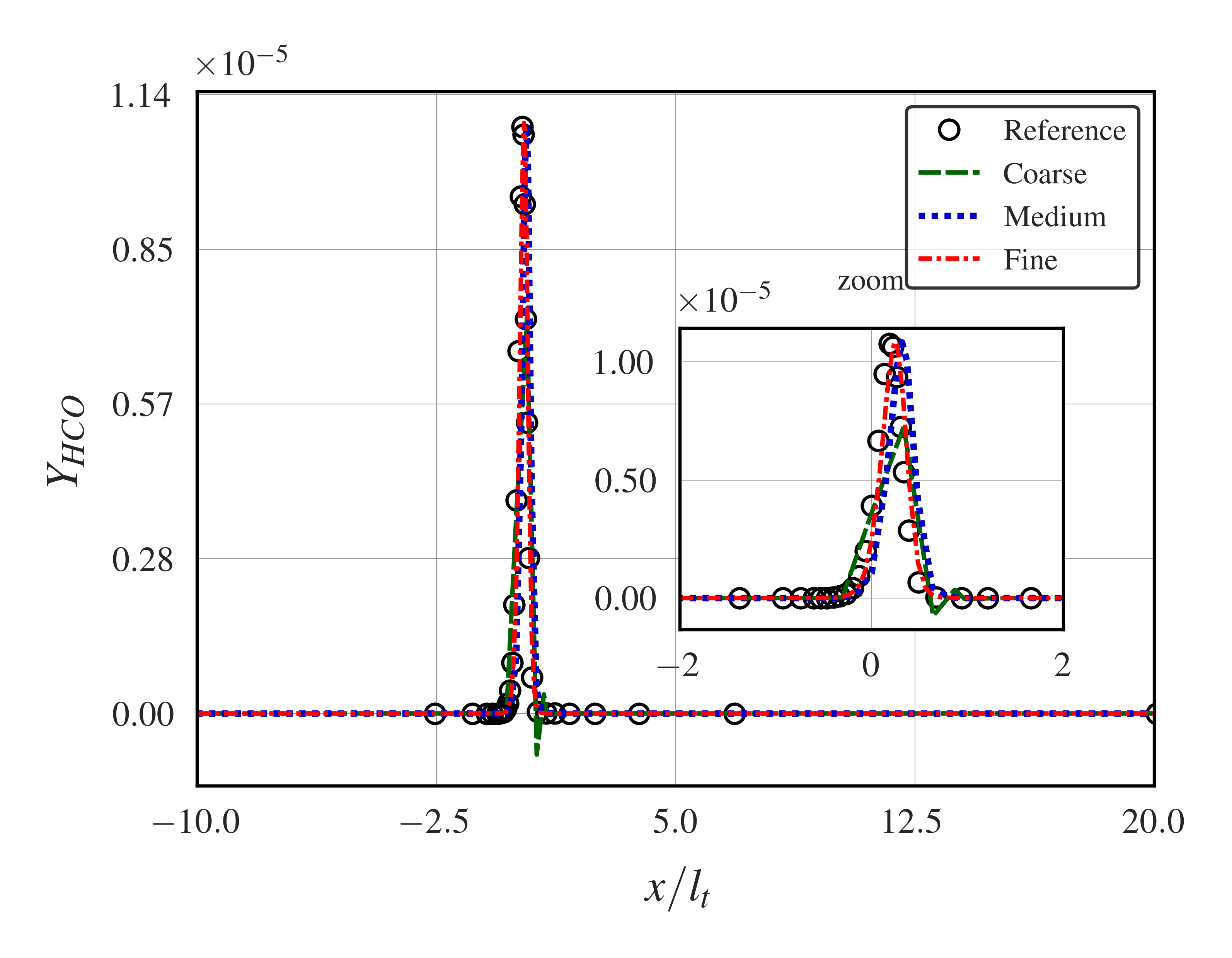}
    \end{subfigure}
    \begin{subfigure}[b]{0.45\columnwidth}
        \includegraphics[width=\columnwidth]{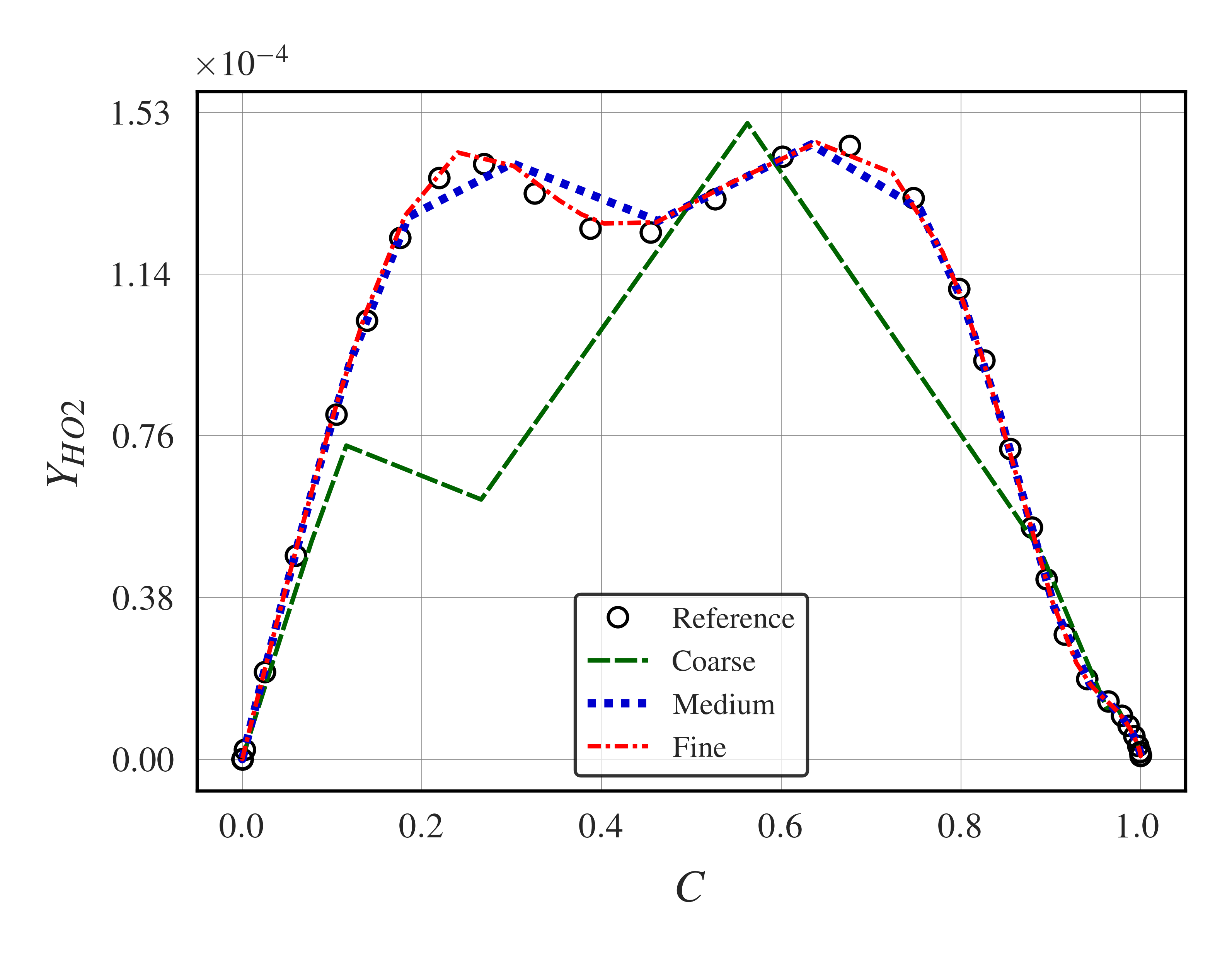}
    \end{subfigure}
    \hfill
    \begin{subfigure}[b]{0.45\columnwidth}
        \includegraphics[width=\columnwidth]{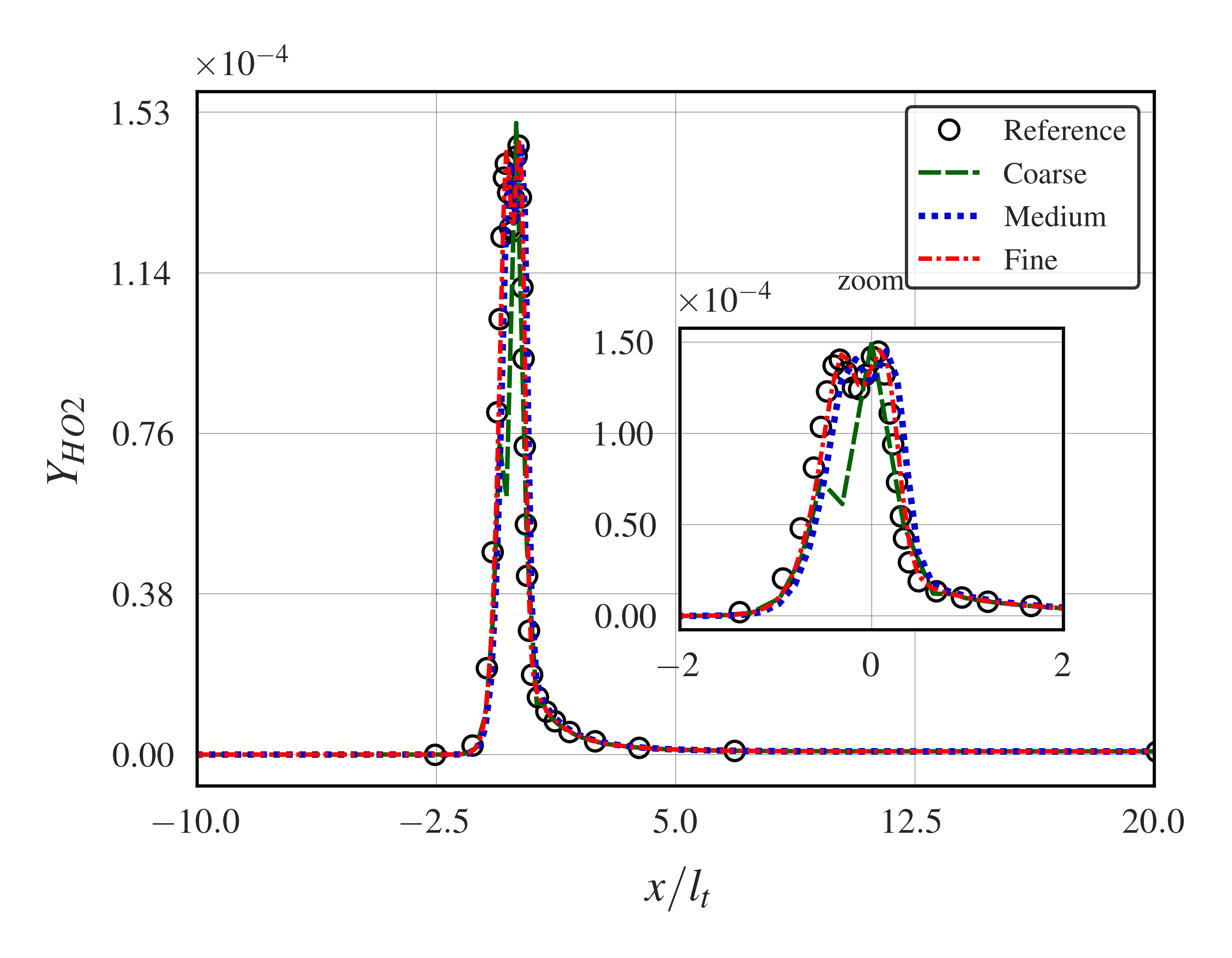}
    \end{subfigure}
    \caption{Comparison of species profiles from DNS results for three grid resolutions: coarse ($73 \mu m$), medium ($48 \mu m$), and fine ($37 \mu m$) against Cantera results. Both progress variables space and physical space are shown.}
    \label{fig: Profiles-val}
\end{figure}

\begin{figure}
    \centering
    \begin{subfigure}{0.45\columnwidth}
    \includegraphics[width=\columnwidth]{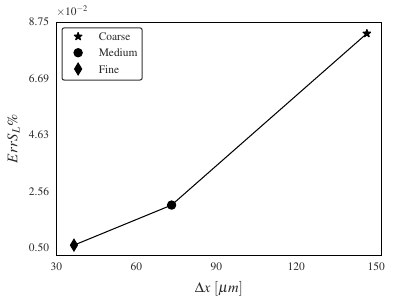}
    \caption{}
    \label{fig: Flamespeed error space}
    \end{subfigure}
    \hfill
    \begin{subfigure}{0.45\columnwidth}
    \includegraphics[width=\columnwidth]{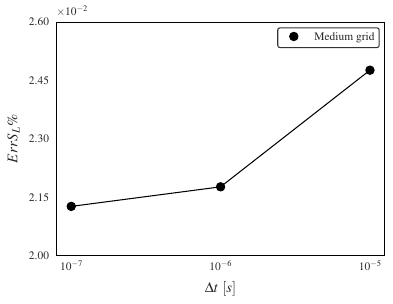}
    \caption{}
    \label{fig: Flamespeed error time}
    \end{subfigure}
    \caption{Deviation of consumption speed, $S_c$, from the laminar burning velocity, $S_L^0$, observed for various spatial resolutions of the grid and time steps.}
    \label{fig: Flamespeed error}
\end{figure}

\figurename~\ref{fig: Profiles-val} displays a comparison between the profiles of two radicals (HCO,HO2) in physical space and progress variable $C$ space as defined in Tab.~\ref{tab: Val-flame}. The reported profiles are obtained using the three different grid resolutions and the Cantera reference solution. The results indicate that the lowest spatial resolution (Coarse grid, resolution $\delta_x=73 \mu m$) cannot accurately reproduce the sharp radical profiles present within the flame front, resulting in significant errors in the flame structure reproduction. Concurrently, it is of fundamental importance to accurately reproduce the flame front propagation speed ($S_L^0$ for planar flames) with minimal error to effectively simulate a premixed flame. The flame speed in the DNS is calculated resorting to a consistent definition of consumption speed~\citep{Poinsot2005,Berger2022_2}:
\begin{equation}
    S_c = \pm \frac{1}{\rho_u Y_{k,u} {L}_y} \int \dot{\omega_k}\ dxdy
\end{equation}
Where $\rho_u$ represents the density of the fresh mixture, $Y_{k,u}$ refers to the unburnt mass fraction of a species that is completely created/depleted through the flame front. Similarly, ${L}_y$ represents the spanwise length of the domain whereas $\dot{\omega_k}$ is the net production/depletion rate of the k-species. The species employed for the evaluation of $S_c$ is the deficient reactant, which is in this case the fuel (methane). Figure~\ref{fig: Flamespeed error}\subref{fig: Flamespeed error space} reports the deviation, defined as $ErrS_L\% = (|S_c-S_L^0|)/S_L^0 \cdot 100$, of $S_c$ obtained by the DNS at the three resolution levels from the reference laminar value $S_L^0$. Overall the error remains limited for all the resolution $\Delta_x$ employed with the medium grid size that represents a good compromise between computational time and accuracy. For this medium resolution, an assessment of the impact of time step $\Delta t$ selection is also performed and reported in Fig.~\ref{fig: Flamespeed error}\subref{fig: Flamespeed error time} where $\Delta t$ is varied from $10^{-5} s$ to $10^{-7} s$. No substantial variation in the flame speed is observed, therefore computational errors remain essentially unaffected by changes in the timestep.

\section{Thermochemical conditions and numerical assessments for ammonia flames}

The target mixture for the present work is 50\%NH3-50\%H2-Air by volume, with an initial temperature of $T_u = 500 K$ and an equivalence ratio of $\Phi = 0.5$. The chemical kinetic mechanism selected for the Ammonia-Hydrogen chemistry is the one developed by~\citet{Gotama2022} featuring 25 species and 110 reactions. Such chemical kinetic mechanism has been proven to be accurate in replicating the laminar burning velocity of hydrogen-enriched ammonia flames, as well as achieving good results on other experimental data such as Ignition delay time (IDT), Jet-stirred reactor (JSR), and flow reactor (FR)~\cite{Szanthoffer2023}. The selected mixture is investigated at atmospheric and elevated pressure and the main flame parameters are reported in Tab.~\ref{tab: NH3-H2-Air-flames}.  

\begin{table}
    \centering
    \begin{adjustbox}{width=0.8\columnwidth}
    \begin{tabular}{c c c c c c}
    \toprule
    Tag  & $p_0$~[Atm] & $\ell_T$~[$\mu$m] & $S_L^0$~[cm/s] & $\tau_F$~[ms] \\
    \midrule
    AH1 & $1$ & $685$ & $34.17$ & $2.00$\\
    \midrule
    AH10 & $10$ & $329$ & $5.44$ & $6.04$\\
    \bottomrule
    \end{tabular}
    \end{adjustbox}
    \caption{Main flame parameters obtained using Cantera of the investigated flames in this study.}
    \label{tab: NH3-H2-Air-flames}
\end{table}

\begin{figure}
    \centering
    \begin{subfigure}[b]{0.3\columnwidth}
        \includegraphics[width=\columnwidth]{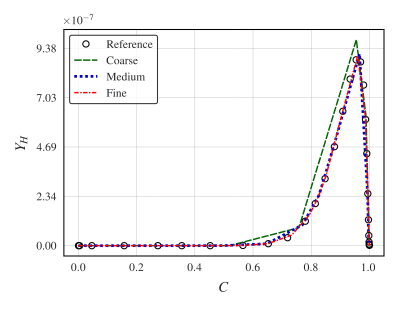}
    \end{subfigure}
    \hfill
    \begin{subfigure}[b]{0.3\columnwidth}
        \includegraphics[width=\columnwidth]{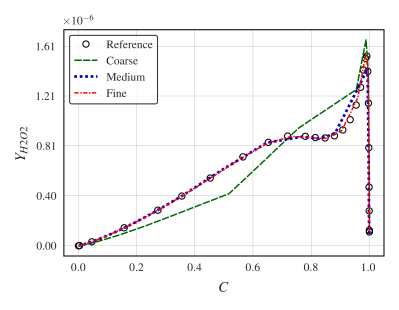}
    \end{subfigure}
    \hfill
    \begin{subfigure}[b]{0.3\columnwidth}
        \includegraphics[width=\columnwidth]{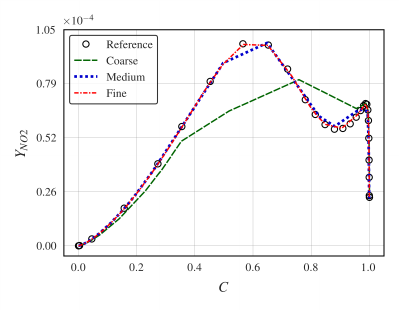}
    \end{subfigure}
    \begin{subfigure}[b]{0.3\columnwidth}
        \includegraphics[width=\columnwidth]{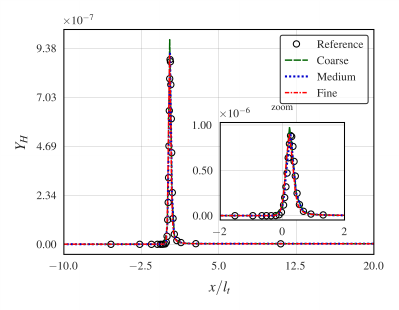}
    \end{subfigure}
    \hfill
    \begin{subfigure}[b]{0.3\columnwidth}
        \includegraphics[width=\columnwidth]{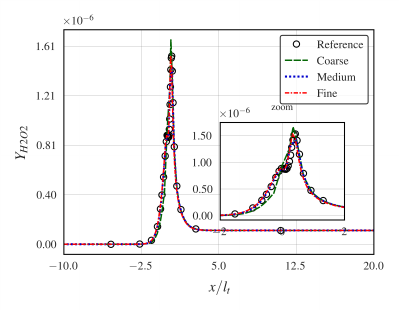}
    \end{subfigure}
    \hfill
    \begin{subfigure}[b]{0.3\columnwidth}
        \includegraphics[width=\columnwidth]{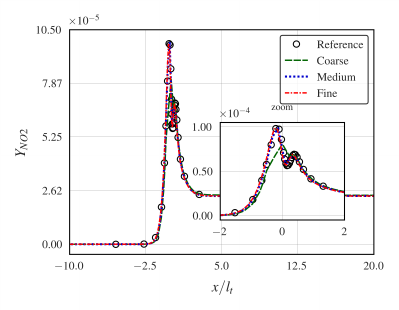}
    \end{subfigure}
    \caption{Comparison of species profiles from DNS results for three grid resolutions: coarse (52 $\mu$m), medium (26$\mu$m), and fine (19$\mu$m) against Cantera results for AH10 flame. Both progress variables space and physical space are shown.}
    \label{fig: Grid_Ammonia}
\end{figure}

To assess the use of a chemical kinetic mechanism for ammonia flames, a similar analysis to the one presented in the previous section is carried out for the AH10 flame. \figurename~\ref{fig: Grid_Ammonia} shows a comparison between the DNS results and the reference Cantera profiles of three representative species (H,H2O2,NO2) in the physical space and the space of progress variables defined, similarly to~\cite{Netzer2021}, as follow:
\begin{equation}
    C = 1-\frac{Y_{H2O}-Y_{H2O}^b}{Y_{H2O}^u-Y_{H2O}^b}
\end{equation}. 
\noindent The profiles shown are obtained using a coarse grid ($\sim8$ points in $\ell_T$), a medium grid ($\sim12$ points in $\ell_T$), and a fine grid ($\sim16$ points in $\ell_T$) while the Cantera solution has been labeled as the reference solution. Overall a good agreement is observable starting from the medium grid in both physical and progress variable space resulting in a flame speed deviation below $\sim1.2\%$. Such resolution is therefore identified as the reference requirement for the simulation of the selected mixture of Tab.~\ref{tab: NH3-H2-Air-flames}.     

The accurate description of transport properties is of fundamental relevance for the combustion of hydrogenated fuels, where the high diffusivity of H and H2 atoms, together with preferential diffusion effects, make predictions of flame propagation characteristics highly sensitive to changes in the diffusion rates of these two species \citep{Sanchez14}. The importance of including thermal diffusion, the Soret effect, in reacting flow simulations is well established in the literature for hydrogen flames. Various studies by \citet{Garcia1984}, \citet{Yang2010}, and Ern and Giovangigli~\citep{Ern98,Ern99} indicate that the Soret diffusion may have a minor impact when dealing with a flat flame that is freely propagating. However, it becomes significant when dealing with a curved or stretched flame, as the cellular structures emerging in the non-linear regime of propagation, for intrinsically unstable flames. In such cases, the thermal diffusion of H2 in the preheat region may affect the amount of fuel that reaches the reaction layer.

\begin{figure}
    \centering
    \begin{subfigure}[b]{0.45\columnwidth}
        \includegraphics[width=\columnwidth]{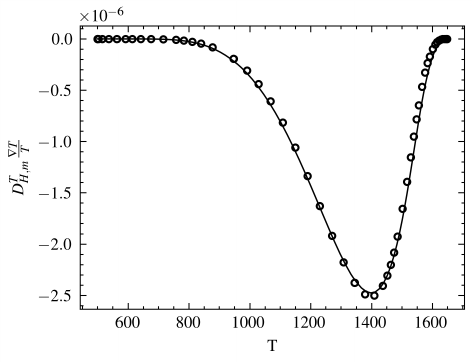}
    \end{subfigure}
    \hfill
    \begin{subfigure}[b]{0.45\columnwidth}
        \includegraphics[width=\columnwidth]{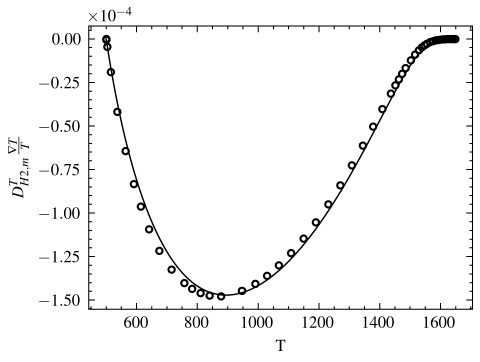}
    \end{subfigure}
    \begin{subfigure}[b]{0.45\columnwidth}
        \includegraphics[width=\columnwidth]{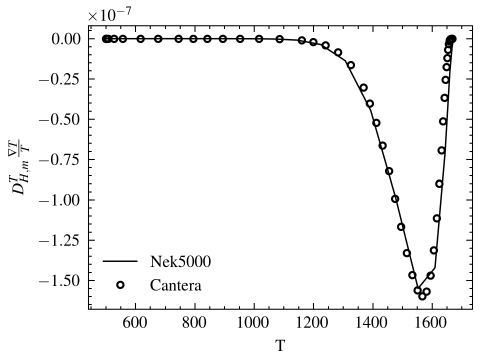}
    \end{subfigure}
    \hfill
    \begin{subfigure}[b]{0.45\columnwidth}
        \includegraphics[width=\columnwidth]{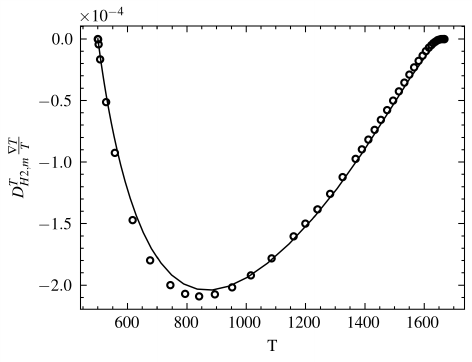}
    \end{subfigure}
    \caption{Comparison of Soret fluxes for H and H2 obtained from DNS with those obtained by the multicomponent model incorporated into Cantera. Top panels: AH1 flame. Bottom panels AH10 flame.}
    \label{fig: Soret_Ammonia}
\end{figure}

In the context of ammonia flames, the role of the Soret effect is still to be completely assessed. For this reason, we performed simulations neglecting and including the thermal diffusion effects in the transport model. From a computational perspective, calculating the thermal diffusion terms presents a challenge. Both the iterative multicomponent method and mixture-averaged thermal diffusion model have a computational cost of $\mathcal{O}(n^2)$ for calculating thermal diffusion coefficients within the chosen chemical model, where $n$ is the number of species. To minimize the computational cost associated with calculating Soret diffusivity, we utilized a model developed by \citet{Schlup2018_1}. This model applies to both molecular and atomic hydrogen diffusivities and scales with $\mathcal{O}(n)$, thereby significantly reducing the local computational effort while maintaining accuracy of thermal diffusion coefficients. As a test on the implemented model, the Soret fluxes for H and H2 obtained from DNS were compared with those obtained by the multicomponent model incorporated into Cantera as shown in Fig.~\ref{fig: Soret_Ammonia} for AH1 and AH10 flames.

\section{Results}
The linear stability of a planar flame is fully characterized by the dispersion relation $\omega(k)$ which represents the growth rate $\omega$ of each small harmonic perturbation of wavenumber $k$. At low wavenumbers, an unstable behavior, $\omega > 0$, is expected due to the prominence of the hydrodynamic DL mechanism which is active at all $k$, while at higher wavenumber (smaller scales) a stabilizing or destabilizing TD effect is expected, depending on the effective Lewis number $Le_{eff}$ being respectively larger or smaller than a critical value $Le_0$. If the TD effects are destabilizing, they will eventually be stabilized at even higher $k$ (yet smaller scales) by transverse heat conduction and reactant diffusion which dampen the perturbation irrespective of the Lewis number~\cite{lapenna2023hydrogen}. The dispersion relation allows for the identification of representative lengthscales related to intrinsic flame instability, such as the cut-off wavelength $\lambda_c$ which is the lengthscale where $\omega=0$ (i.e. at which hydrodynamic and diffusive effects balance) and the lengthscale of maximum growth rate $\lambda_{\omega_{max}}$. The former is important to determine if a flame in a particular domain will experience the onset of IFI corrugations~\cite{lapenna2019large} while the latter is related to the most probable cellular-wrinkle size that will emerge in the non-linear regime of propagation~\cite{Berger2019}. Such lengthscales are intrinsic properties of the flame as they depend only on the thermochemical conditions, namely the pressure $p$, the fresh gas temperature $T_u$, the equivalence ratio $\phi$, and fuel type and composition. The dispersion relation can be evaluated by both resorting to theoretical models and by performing DNS. In this section, we first discuss the qualitative outcomes of a theoretical model. Then we perform DNS to quantitatively investigate the role of pressure and Soret effects on the stability limits of the target hydrogen-enriched ammonia/air flame. The calculation of dispersion relations through DNS allows no assumptions to be made about the thermochemical properties of the mixture.

\subsection{Theoretical stability limits}
The theoretical results of Matalon et al.~\cite{Matalon2003} are employed in this work as also done in other recent DNS works on pure hydrogen/air flames~\cite{Berger2022_2,howarth2022empirical}. A detailed discussion on the various forms of the model dispersion relation is given in~\cite{Matalon2007}. The model dispersion relation reads~\cite{Matalon2003}: 

\begin{equation}
    \tilde{\omega} = \omega_{DL} \tilde{\kappa} + \omega_2 \tilde{\kappa^2}
    \label{Eq: Grate}
\end{equation} 
where $\tilde{\omega} =\omega \tau_f$, while $\tilde{\kappa} =\kappa \ell_T$ and he Darrieus-Landau coefficient $\omega_{DL}$ takes the form:
\begin{equation}
     \omega_{DL} =\frac{1}{\sigma +1} \sqrt{\sigma^3 +\sigma^2 -\sigma} -\sigma
     \label{Eq: wdl}
\end{equation}
where $\sigma= \rho_u/\rho_b$ is the thermal expansion ratio. Note that the model is limited to $\mathcal{O}(k^2)$ terms and does not incorporate higher  $\mathcal{O}(k^4)$ stabilizing terms~\cite{lapenna2023hydrogen}.
All the other thermochemical parameters of the flame, namely Prandtl $Pr$, Zel'dovich $Ze$, and $Le_{eff}$, are included in the $\omega_{2}$ diffusive coefficient. This coefficient is negative when TD effects are stabilizing and vice versa, and reads:     
\begin{equation}
     \omega_{2} = B_1 + Ze(Le_{eff} -1)B_2 + Pr B_3
\end{equation}
where the parameters $B1$, $B2$, $B3$, are a function of $\sigma$ and can incorporate a dependence on temperature of transport coefficients ~\cite{Matalon2003}. The flame parameters to estimate $\omega_2$ are calculated as follows, starting from the Zel'dovich number:    
\begin{equation}
    Ze = \frac{E}{R} \frac{\left( T_u-T_b \right)}{T_u^2}
\end{equation}
where $R$ is the universal gas constant, $T_b$ is the adiabatic flame temperature and $E$ is the global activation energy estimated, following~\cite{Law2000} as: $E = R \frac{2d(\rho_u S_L^0)}{d(1/T_u)}$ by means of a set of 1D unstretched premixed flame calculations. The mixture's effective Lewis number is evaluated following~\cite{Matalon2003}:
\begin{equation}
    Le_{eff}=\frac{Le_O+ALe_F}{1+A} 
\end{equation} 
being $A$ defined for lean mixtures as $A=1+Ze(\phi^{-1} -1)$ while $Le_O$ is the oxidizer Lewis number and $Le_F$ is the fuel mixture Lewis number evaluated using the volume-based formulation reported in~\cite{bouvet2013effective}. In this framework, for a particular $\sigma$ and $Ze$ a critical Lewis number can be evaluated:
\begin{equation}
    Le_0 =1-\frac{B_1+Pr B_3}{\beta B_2}
\end{equation}
as it is strictly related to the sign of $\omega_2$,where positive values of $\omega_2$ are obtained when $Le_{eff}<Le_0$. In this case, however, the model dispersion relation will diverge lacking higher-order stabilizing terms.

\begin{figure}
    \centering
    \begin{subfigure}[b]{0.48\columnwidth}
        \includegraphics[width=\columnwidth]{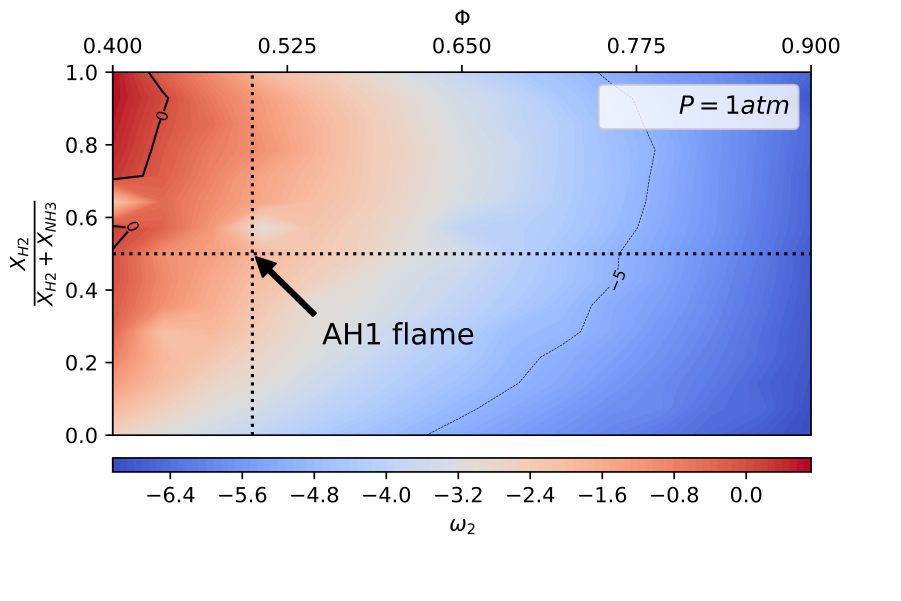}
    \end{subfigure}
    \hfill
    \begin{subfigure}[b]{0.48\columnwidth}
        \includegraphics[width=\columnwidth]{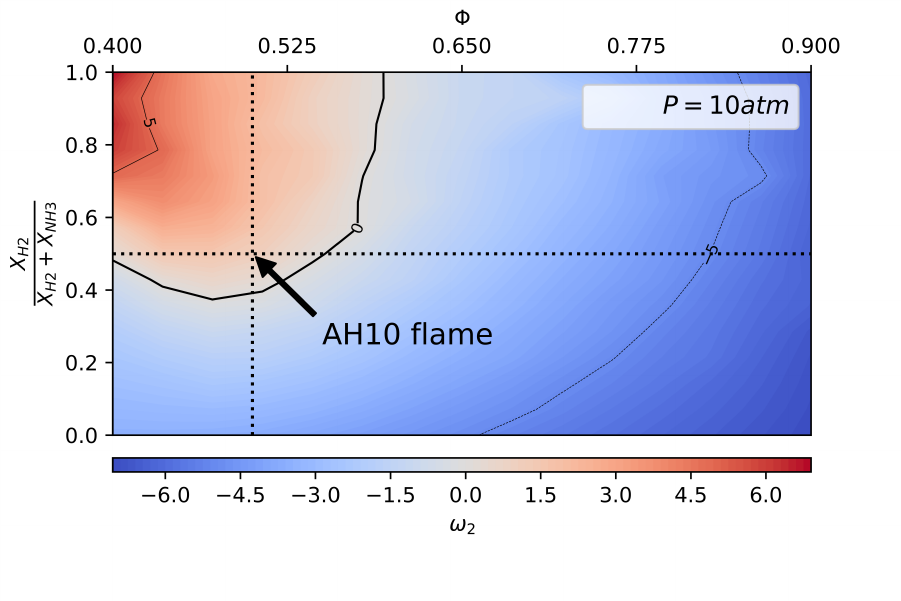}
    \end{subfigure}
    \caption{$\omega2$ field as a function of equivalence ratio, and hydrogen volumetric percentage in the mixture for AH1 and AH10 flames.}
    \label{fig: w2-map}
\end{figure}

Using the flame parameters definition previously reported, it is now of interest to evaluate the $\omega_2$ coefficient for hydrogen-enriched ammonia flames and use it as an approximate criterion to establish the mixtures that are expected to be TD-unstable. \figurename~\ref{fig: w2-map} shows $\omega2$ as a function of $\phi$ and hydrogen content in ammonia mixtures, at the two target pressures selected for this work. As expected, higher and positive values of $\omega_2$ can be observed for leaner mixtures and higher hydrogen contents. The conditions of the AH1 and AH10 flames are also reported and it can be observed that the AH1 flame is expected to be TD-stable ($\omega_2<0$) while the AH10 TD-unstable ($\omega_2>0$). It is worth mentioning that this criterion is as accurate as the simplifying assumptions used in the context of the hydrodynamic model used to derive Eq.\ref{Eq: Grate}.
This further motives the linear stability analysis performed using DNS for the two target flames, as described in the next section. For completeness, all the flame parameters for the AH1 and AH10 flame are summarized in Tab.~\ref{tab: Properties} where it can be noted that $Le_{eff}>Le_0$ for AH1 and $Le_{eff}<Le_0$ for AH10. 

\begin{table} 
    \centering
    \begin{adjustbox}{width=.8\columnwidth}
    \begin{tabular}{c c c c c c c c c}
    \toprule
    Tag  & $Ze$[-] & $Le_e$[-] & $Le^*$[-] & $B_1$[-] & $B_2$[-] & $B_3$[-] & $\omega_{dl}$[-] & $\omega_2$[-]\\
    \midrule
    AH1 & $8.14$ & $0.76$ & $0.59$ & $4.77$ & $1.57$ & $0.68$ & $0.75$ & $-2.10$\\ 
    \midrule
    AH10 & $14.26$ & $0.74$ & $0.77$ & $4.83$ & $1.59$ & $0.70$ & $0.76$ & $0.60$\\ 
    \bottomrule
    \end{tabular}
    \end{adjustbox}
    \caption{Summary of the key parameters required for calculating the theoretical dispersion relation through the Matalon-Bechtold theoretical model.}
    \label{tab: Properties}
\end{table}

\subsection{Numerical dispersion relations}
\begin{figure}
    \centering
        \includegraphics[width=0.6\columnwidth]{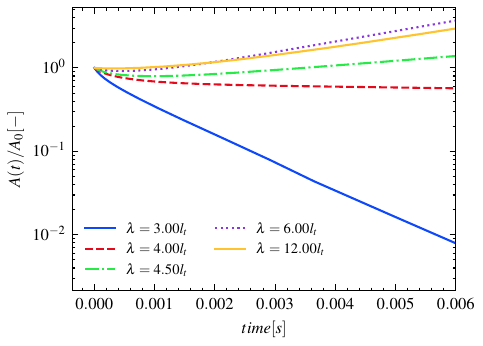}
    \caption{Time evolution of the normalized perturbation amplitude $A(t)/A_0$ for different perturbation wavelengths $\lambda$ in the linear regime for AH1 flame.}
    \label{fig: Transition}
\end{figure}

To calculate the growth rates, multiple DNS simulations are conducted with a planar flame that is selectively perturbed with a single wavelength harmonic perturbation. The flame is initialized within the planar 2D configuration with a 1D unstrected laminar flame and the perturbation is imposed on the flame position. The amplitude of the chosen initial perturbation $A_0$ is chosen as $8\%$ of the thermal flame thickness $\ell_T$ to remain within the assumption of small perturbations. The length of the domain in the propagation direction remains constant throughout the linear stability analysis at $L_x=50\ell_T$ for all DNSs to ensure that the zero gradient outlet condition is well posed behind the flames without any influence on the flame development. On the other hand, the domain dimension $L_y$ in the lateral spanwise direction is changed for each DNS to accommodate perturbations at different wavelengths $\lambda = L_y$ with periodic boundary conditions. To ensure a proper resolution of the small perturbation imposed, a uniform grid is employed featuring $\sim50$ grid points resolving the thermal flame thickness. The amplitude of the perturbation is subsequently tracked over time $A(t)$ and the growth rate is calculated, as done in previous studies~\cite{Attili2021,lamioni2020pressure,Berger2022_1}, as:
\begin{equation}
    \omega= \frac{d\log{A(t)}}{dt}
\end{equation}
\noindent The time evolution of the amplitude of the perturbations are shown in Fig.~\ref{fig: Transition} for the AH1 flame. The simulations featuring long wavelength perturbations have exponentially increasing amplitudes $A(t)$, while smaller wavelength perturbations have exponentially decreasing amplitudes. The growth rate is then deduced from the slope of the amplitude evolution in time and and the numerical dispersion relation is obtained. 

\begin{figure}
    \centering
    \begin{subfigure}[b]{0.49\columnwidth}
        \includegraphics[width=\columnwidth]{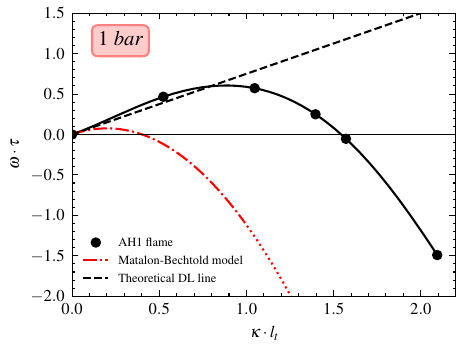}
        \caption{}
    \end{subfigure}
    \hfill
    \begin{subfigure}[b]{0.49\columnwidth}
        \includegraphics[width=\columnwidth]{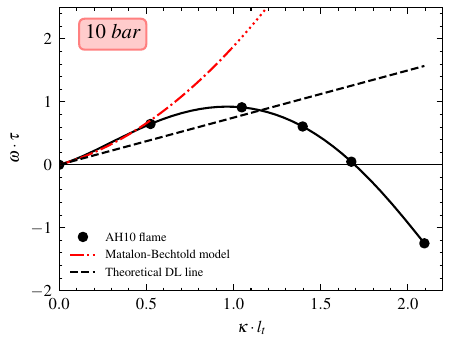}
        \caption{}
    \end{subfigure}
    \caption{Comparison of the numerically computed dispersion relation with the model dispersion relation for AH1 and AH10 flames. 
    }
    \label{fig: Dispersion_teoretical}
\end{figure}

\figurename~\ref{fig: Dispersion_teoretical} compares the numerical dispersion relations to the model dispersion relations for the AH1 and AH10 flames. As expected, a significant discrepancy can be observed due to the simplifying assumptions of the theoretical model. Moreover, the lack of a higher-order stabilizing term, which is expected to become dominant for $k>1$, causes the theoretical dispersion relation to diverge for the AH10 flame which is expected to be TD unstable.
In this latter case, DNS simulations are the only viable tool to construct a complete dispersion relation.

\begin{figure}
    \centering
    \begin{subfigure}[b]{0.49\columnwidth}
        \includegraphics[width=\columnwidth]{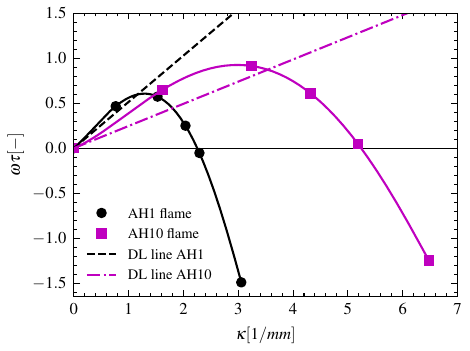}
        \caption{}
    \end{subfigure}
    \hfill
    \begin{subfigure}[b]{0.49\columnwidth}
        \includegraphics[width=\columnwidth]{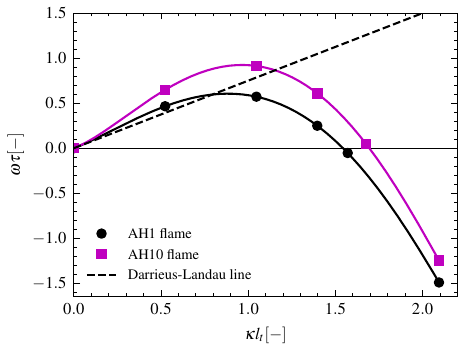}
        \caption{}
    \end{subfigure}
    \caption{Numerically computed dispersion relations for different pressure values corresponding to flames in Tab.~\ref{tab: NH3-H2-Air-flames}. (\textbf{a}): Dimensional dispersion relation. (\textbf{b}): Dimensionless dispersion relation. }
    \label{fig: Dispersion_pressure}
\end{figure}

\figurename~\ref{fig: Dispersion_pressure} compares the numerical dispersion relations for the AH1 and AH10 flames as a function of both a dimensional and non-dimensional wavenumber. The symbols represent the values of $\omega$ obtained from each simulation while the fitted curves are obtained using shape interpolant splines. The linear DL term of Eq.~\ref{Eq: wdl} is also reported for reference. The higher-pressure flame AH10 features a wider range of unstable wavelengths as well as higher non-dimensional growth rates. Interestingly, both cases exhibit growth rates that exceed those associated with the hydrodynamic instability mechanism $\omega_{DL}$, indicating a positive contribution from the TD mechanism. While this was to be expected for the AH10 flame, the theoretical model and the ensuing definition of the critical Lewis number suggested the opposite for AH1. Overall, the increased pressure reduces the critical wavelength dimensions $\lambda_c$ (higher wavenumber $\kappa_c$) while the growth rate $\omega$ linked with the unstable wavelengths is increased as pressure rises, consistently with other results obtained for hydrogen-enriched ammonia flames~\cite{Gaucherand2023} and other mixtures~\cite{Attili2021,Berger2022_1}.

\begin{figure}
    \centering
    \begin{subfigure}[b]{0.49\columnwidth}
        \includegraphics[width=\columnwidth]{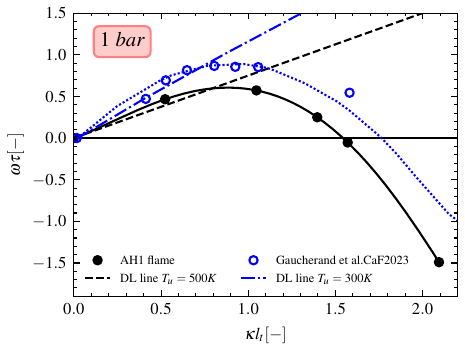}
        \caption{}
    \end{subfigure}
    \hfill
    \begin{subfigure}[b]{0.49\columnwidth}
        \includegraphics[width=\columnwidth]{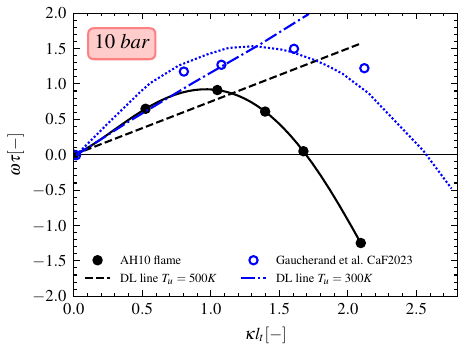}
        \caption{}
    \end{subfigure}
    \caption{Effect of $T_u$ on dispersion relations for different pressures. Data at $T_u =300K$ (blue lines and void markers) taken from \citep{Gaucherand2023}.}
    \label{fig: Dispersion_temperature}
\end{figure}

Figure~\ref{fig: Dispersion_temperature} illustrates the impact of temperature on dispersion relations at the two pressures investigated. We compared the dispersion relation obtained for AH1 and AH10 flames using a fresh mixture temperature of $T_u=500K$ with the dispersion relation obtained for flames at the same composition, equivalence ratio, and pressure, but a different $T_u$ at $=300K$ calculated by \citet{Gaucherand2023}. Consistent with findings by \citet{Berger2022_1} for pure hydrogen flames, elevating the temperature of the unburnt mixture leads to an increased critical wavelength, resulting in a stabilizing effect at both the pressures investigated. 

\begin{figure}
    \centering
    \includegraphics[width=0.8\columnwidth]{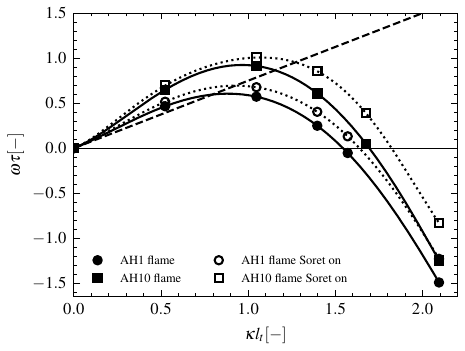}
    \caption{Effect of the Soret diffusion on the dispersion relations at different pressures.}
    \label{fig: Dispersion_soret}
\end{figure}

\begin{table} 
    \centering
    \begin{adjustbox}{width=.7\columnwidth}
    \begin{tabular}{c c c c c c c c c}
    \toprule
    Flame  &$\lambda_c [mm]$ & $\lambda_c^s [mm]$ & $\lambda_{\omega_{Max}} [mm]$ & $\lambda_{\omega_{Max}}^s [mm]$\\
    \midrule
    AH1 & $2.79$ & $2.63$ & $4.87$ & $4.65$\\
    \midrule
    AH10 & $1.19$ & $1.11$ & $2.11$ & $1.90$\\
    \bottomrule
    \end{tabular}
    \end{adjustbox}
    \caption{Key points in stability limits for AH1 and AH10 flames. Superscript "S" denotes results with the Soret effect activated.}
    \label{tab: Results}
\end{table}

Finally, the impact of thermophoresis (Soret effect) on flame stability is evaluated by including it in the governing equation and calculating two additional dispersion relations at the conditions of AH1 and AH10 flames. The comparison between the dispersion relations evaluated with and without the Soret effect is shown in Fig.~\ref{fig: Dispersion_soret} using non-dimensional units.
The Soret effect is shown to affect the dispersion relations, the ensuing cut-off wavelengths, and the maximum growth rates. For the atmospheric pressure case AH1 flames, including the Soret effect causes $\lambda_c$ to decrease from $2.79\ mm$ to $2.63\ mm$ , while the maximum growth rate is also increased from $302\ s^{-1}$ to $339\ s^{-1}$, i.e. a $5.7\%$ decrease in the $\lambda_c$  and a $9\%$ increase of $\omega_{max}$. Similarly for the AH10 flame, including the Soret effect in the transport models resulted in a decrease of $\lambda_c$ from $1.19\ mm$ to $1.11\ mm$, while $\omega_{max}$ increased from $153\ s^{-1}$ to $167\ s^{-1}$, i.e. a $6.7\%$ decrease for $\lambda_c$ and a $9.2\%$ increase for $\omega_{max}$. Overall, the dispersion relations indicate that the effect of the extra diffusion induced by thermophoresis affects the flame stability, leading to a flame that is more prone to feature the onset of IFI. Finally, in Tab.~\ref{tab: Results} we reported the key results of our investigation on the stability limits of AH1 and AH10 flames. 

\section{Conclusion}
In this work, we presented linear stability analyses for hydrogen-enriched ammonia/air flames (50\%H2-50\%NH3 by volume) using direct numerical simulations with a detailed chemical kinetic mechanism and transport. An in-house modified version of the high-order, spectral element, solver \textit{nek5000} has been validated to perform such DNS. It has been demonstrated that our numerical framework is capable of accurately reproducing the flame structure of laminar premixed flames using typical DNS-type grid sizes and with a small error in the flame speed compared to Cantera. The thermophoresis effect has also been included and validated using a computationally efficient model from the literature. 

The stability limits of an NH3/H2/air mixture at both atmospheric pressure and 10 atm were investigated qualitatively through existing theoretical models in the literature, and subsequently through DNS. The numerical dispersion relations, when compared to the model dispersion relations reveal that the latter have limited quantitative predictive capabilities. On the other hand, the existing theoretical framework can have an important qualitative role in establishing potential TD instability regions in the space of parameters such as equivalence ratio and hydrogen content.
The impact of pressure and the Soret effect have been assessed by comparing the numerical dispersion relations. The ensuing data indicates that both pressure and the Soret effect promote the onset of intrinsic instabilities. The pressure is shown to have a similar impact as observed with other mixtures in the literature such as lean methane/air and hydrogen/air premixed flames. It is found that the Soret effect influences the stability limits of flames by reducing the critical wavelength $\lambda_c$ and increasing the maximum growth rate $\omega_{Max}$. However, neglecting the supplementary diffusion effects originating from the Soret effect leads to inaccuracies on the order $\sim5\%$ on the stability limits of the investigated flames.  

\section*{Declaration of Competing Interests}
The authors declare that they have no known competing financial interests or personal relationships that could have appeared
to influence the work reported in this paper.

\section*{Acknowledgement}
This work has been supported by Baker-Hughes and Lazio Region. P.E.L acknowledges the support of Sapienza University for the early-stage researchers' funding for the project “A pragmatic support to the hydrogen economy: data-driven modeling of high-pressure combustion for propulsion and power” and for the small-size research project "Combustion under extreme thermodynamic conditions for green propulsion and power". The Italian supercomputing center CINECA is acknowledged for the award, under the ISCRA initiative, of high-performance computing resources and support for the project \textit{IsB26-Hydrogen}. This work has been also supported by ICSC (Centro Nazionale di Ricerca in High-Performance Computing, Big Data and Quantum Computing) funded by the European Union – NextGenerationEU.

\bibliographystyle{elsarticle-num-names} 
\bibliography{Bibliography}





\end{document}